\documentclass{article}

\usepackage{arxiv}

\usepackage[utf8]{inputenc} 
\usepackage[T1]{fontenc}    
\usepackage{hyperref}       
\usepackage{url}            
\usepackage{booktabs}       
\usepackage{amsfonts}       
\usepackage{nicefrac}       
\usepackage{microtype}      
\usepackage{lipsum}
\usepackage{cite}
\pdfoutput=1

\usepackage{graphicx}
\graphicspath{{./Figures/}}
\DeclareGraphicsExtensions{.pdf,.jpg,.png}

\usepackage{mathrsfs,amsmath}
\usepackage{mathtools}
\usepackage{algorithm}
\usepackage{algpseudocode}

\usepackage{amsfonts}
\usepackage{tabularx}
\usepackage{multicol}
\usepackage{multirow}
\usepackage{array}
\usepackage{enumitem}
\usepackage{placeins}
\usepackage{hhline}
\usepackage[bottom]{footmisc}
\usepackage{soul,xcolor}
\usepackage{float}
\usepackage{textcomp}

\title{Speckle illumination spatial frequency domain imaging for projector-free optical property mapping}

\author{
  Mason T.~Chen \\
    Department of Biomedical Engineering\\
    Johns Hopkins University\\
    3400 N. Charles Street, Baltimore, MD, 21218, USA \\
   \And
  Melina Papadakis \\
    Department of Biomedical Engineering\\
    Johns Hopkins University\\
    3400 N. Charles Street, Baltimore, MD, 21218, USA \\
    \And
  Nicholas J. Durr \\
    Department of Biomedical Engineering\\
    Johns Hopkins University\\
    3400 N. Charles Street, Baltimore, MD, 21218, USA\\
  \texttt{ndurr@jhu.edu}
}

\begin{document}
\maketitle

\begin{abstract}
Spatial Frequency Domain Imaging can map tissue scattering and absorption properties over a wide field of view, making it useful for clinical applications such as wound assessment and surgical guidance. This technique has previously required the projection of fully-characterized illumination patterns. Here, we show that random and unknown speckle illumination can be used to sample the modulation transfer function of tissues at known spatial frequencies, allowing the quantitative mapping of optical properties with simple laser diode illumination. We compute low- and high-spatial frequency response parameters from the local power spectral density for each pixel and use a look-up-table to accurately estimate absorption and scattering coefficients in tissue phantoms, \textit{in-vivo} human hand, and \textit{ex-vivo} swine esophagus. Because speckle patterns can be generated over a large depth of field and field of view with simple coherent illumination, this approach may enable optical property mapping in new form-factors and applications, including endoscopy. 
\end{abstract}

\maketitle

Tissue optical properties, especially the absorption ($\mu_{a}$) and reduced scattering coefficients ($\mu'_{s}$), provide useful information on tissue composition, oxygenation, and metabolism \cite{WEINKAUF2019555,mourant2000light,Steelman:19}. Absorption and scattering maps are being used for an increasing variety of clinical applications, including image-guided surgery, wound monitoring, and assessment of surgical margins \cite{nguyen2013novel,kaiser_noninvasive_2011,maloney2018review}. 
Moreover, unlike color values, optical properties are absolute measurements that can be directly compared across different imaging platforms, study sites, and time-scales, facilitating their statistical interpretation via machine learning. 

Over the last decade, Spatial Frequency Domain Imaging (SFDI) has emerged as a powerful tool for optical property measurements. Conventional SFDI acquires images of tissue under sinusoidal illumination at different spatial frequencies and phase offsets, sampling the modulation transfer function (MTF) of the tissue \cite{cuccia_quantitation_2009}. Because the bulk tissue scattering and absorption properties preferentially attenuate the high- and low-spatial frequencies of the MTF, respectively, sampling the MTF at a range of spatial frequencies allows decoupling the two effects. Images are subsequently demodulated and calibrated, and model inversion is performed using a lookup table (LUT). SFDI can rapidly generate wide-field optical property maps using a camera and projector in a non-contact configuration. These advantages have led to the exploration of SFDI for a number of clinical and research applications \cite{nguyen2013novel,kaiser_noninvasive_2011,yafi2017quantitative,tabassum2016feasibility,WEINKAUF2019555,maloney2018review}. 

Despite its advantages, there are several practical challenges to translating SFDI to clinical settings. To generate an optical property map, conventional SFDI requires a minimum of six images at each wavelength (three phase offsets at two different spatial frequencies) and performs a pixel-wise LUT search. Progress has been made towards improving the acquisition and processing time. For example, the acquisition requirements can be relaxed by estimating the MTF from a single spatial frequency using signal processing \cite{vervandier_single_2013} and content-aware machine learning \cite{8943974}. Processing speed has also been improved with GPU-based implementations and machine learning \cite{zhao2018deep,aguenounon2020real,chen2020rapid}. A second major challenge is that SFDI requires the projection of structured illumination with precisely controlled imaging geometry and known spatial frequencies. This requirement has made the translation of SFDI to endoscopy and other space-constrained applications particularly difficult. Previous work has implemented SFDI in custom benchtop endoscopy systems with an added projection channel \cite{angelo_real-time_2017,kress2017dual}, but this approach requires either rigid relay optics or low-pixel-count fiber imaging bundles, each of which take up significant cross-sectional area of the endoscope.

\setlength{\belowcaptionskip}{-7pt}
\begin{figure}[tb]
\centering
\includegraphics[width=\linewidth]{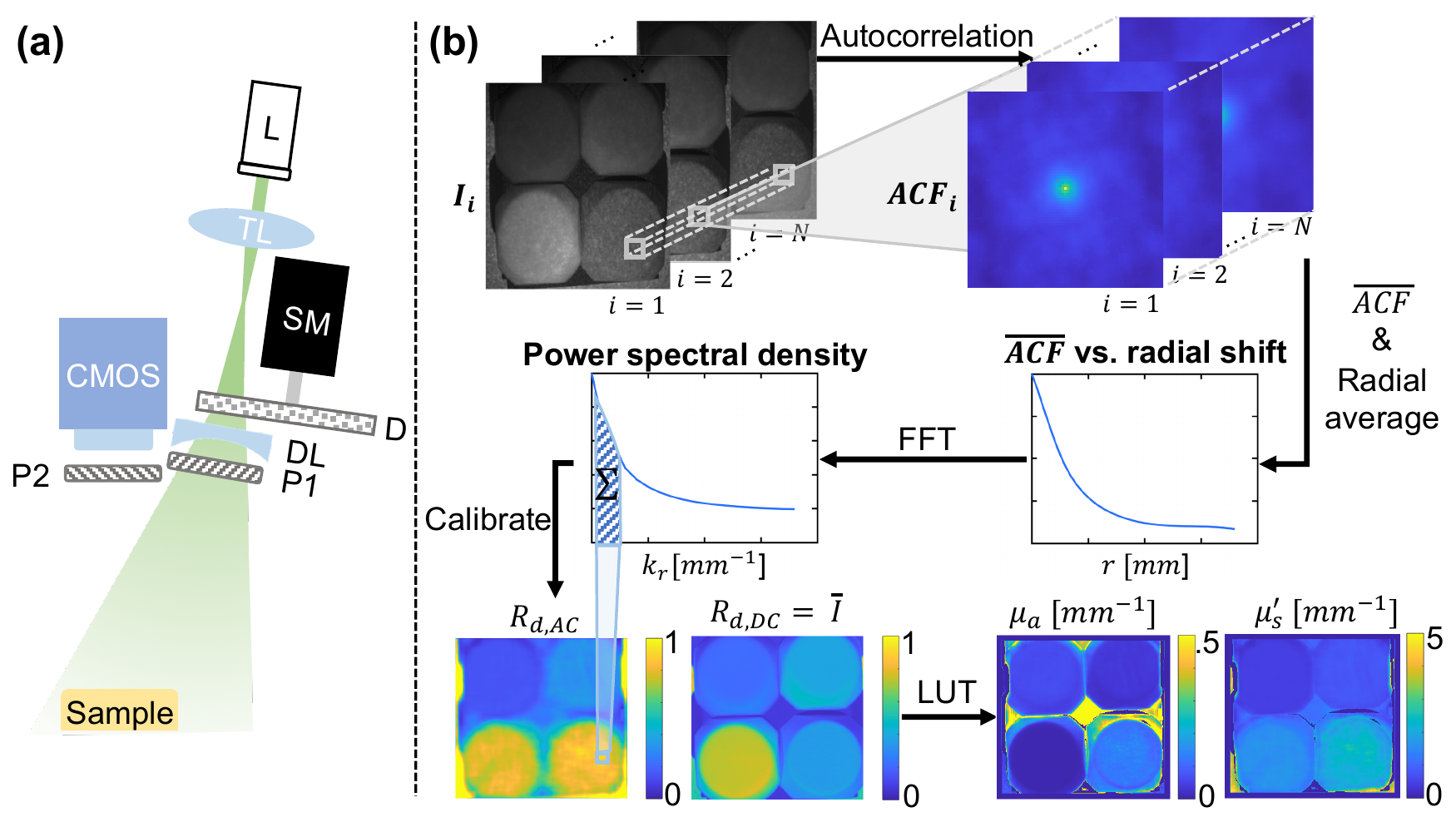}
\caption{Overview of si-SFDI. (a) Experimental setup. A laser diode (L) is focused by a tunable lens (TL) on a rotating diffuser (D) mounted on a stepper motor (SM). A diverging lens (DL) spreads the light to match imaging vergence. Two linear cross polarizers (P1 and P2) reduce specular reflections. (b) Processing flow. The mean autocorrelation function (ACF) is calculated on a sliding window for $N$ speckle images. Results are radially averaged and the Fast Fourier Transform (FFT) is taken to produce a local power spectral density (PSD). After calibration, AC and DC reflectances ($R_{d,AC}$ and $R_{d,DC}$) are used to estimate optical properties from a lookup table (LUT).}
\label{fig:overview}
\end{figure}

An alternative to projecting images of patterns to sample the MTF is to utilize the speckle patterns formed by the interference of coherent illumination. Analyzing tissue response to speckle patterns is often used to estimate blood flow in laser speckle contrast imaging \cite{dunn2001dynamic,boas2010laser}. This approach has the important advantage of being amenable to low-cost and compact implementation, for example, by laser illumination coupled through an optical fiber. Moreover, compared to image projection, coherent illumination can produce patterns with high spatial frequencies over a large depth of field. Recently, Jain et al. \cite{jain2019measuring} analyzed the response of turbid liquid phantoms to laser illumination and found that speckle patterns are blurred in agreement with a model based on the optical properties of the sample. However, this technique has several limitations. First, this work developed a forward model that correlates with sample measurements but did not directly measure optical properties or evaluate accuracy. Second, this model analyzes full images, producing one response measurement for an entire image, and thus is not capable of producing an optical property map and only works on homogeneous samples.

In this study, we present speckle illumination spatial frequency domain imaging (si-SFDI), which maps the optical properties of turbid media from unknown laser speckle patterns. For the same imaging geometry and optical properties, the integral of the power spectral density (PSD) should be constant for any speckle pattern randomization. Therefore, for the same imaging geometry but different optical properties, relative changes in the MTF can be sampled at consistent spatial frequencies via a phantom calibration, but without knowledge of the exact speckle pattern. This phenomenon can be exploited to reconstruct tissue optical property maps using as few as one speckle pattern. There are two main contributions of this study. First, we develop an inverse model to accurately measure tissue optical properties from random speckle images. Second, we apply this technique to heterogeneous, biological tissues and evaluate its performance. To our knowledge, this is the first attempt to directly map optical properties from unknown laser speckle patterns.

The experimental setup is shown in Fig. \ref{fig:overview}(a). A 520 nm wavelength laser diode is used for illumination (Opt Lasers micro RGB laser module) and a 1392 x 1040 pixel CMOS sensor with a camera lens ($f/\#$ = 2.1) is used for imaging, resulting in 138 $\mu$m pixels in the object space when the sample is 30 cm away. A tunable lens (Optotune EL-10-30-TC) focuses the laser beam onto a 220-grit diffuser (Thorlabs DG20-220). By varying focal length, the tunable lens alters the laser spot size on the diffuser surface and subsequently, the size of the speckles in the object space. Speckle size was adjusted to be an average of 3 pixels per speckle grain width to satisfy the Nyquist criterion. The diffuser is mounted on a NEMA 11 stepper motor, which rotates at random angles to vary the speckle pattern. The light is spread by a diverging lens (Thorlabs LD1613-A) to match the 60\textdegree camera field of view. Additionally, the illumination and detection paths are cross-polarized to reduce specular reflections. The laser illumination path is mounted at a 30-degree angle on the side of a commercial SFDI imaging head (Modulim Reflect RS\textsuperscript{TM}), which has a projector at a 12\textdegree  angle and was used for measuring ground truth optical properties at 526 nm using 3-phase 0 and 0.2$mm^{-1}$ sinusoidal illumination.

Figure \ref{fig:overview}(b) summarizes the si-SFDI algorithm. Laser speckle illumination, which is randomized by object and diffuser movements, is a wide-sense stationary process \cite{jain2019measuring,goldfischer1965autocorrelation,goodman2007speckle} with a constant autocorrelation function (ACF) and power spectral density (PSD). We characterize the tissue response in the spatial frequency domain by analyzing the PSD of each speckle image. This allows tissue optical properties to be calculated without knowledge of the exact illumination pattern. The image of the remitted signal, $v(x, y)$, can be expressed as the convolution of the impulse response $h(x, y)$ of the sample and system with the illumination speckle pattern $u(x, y)$, or $v(x, y) = h(x, y) * u(x, y)$. To reduce computational cost, we estimate the ACF utilizing the Wiener-Khinchin theorem:

\begin{equation}
a_v(x, y) = \mathcal{F}^{-1}[V^*(k_x, k_y) \cdot V(k_x, k_y)],
\label{eq:wiener}
\end{equation}
where $V$ and $V^*$ represent the Fourier transform of the output image $v$ and its conjugate. Taking the magnitude of the Fourier transform of the $a_v$, we obtain the power spectral density $S_V(k_x, k_y)$. Assuming the impulse response is radially symmetrical, we can express the power spectral density (PSD) as:

\begin{equation}
S_V(k_r) = |H(k_r)|^2 \cdot S_U(k_r).
\label{eq:psd}
\end{equation}
Thus, given a measured image of a reference phantom with known optical properties under coherent illumination, the ratio between the PSDs can be calculated and subsequently scaled by the reference model response predicted by Monte Carlo simulations. In this way, we characterize the sample response $H(k_r)$ without measuring the input pattern $u(x, y)$.

In order to reduce noise in the 0-lag PSD, we apply a low-pass median filter with a kernel size of 3 pixels to the raw image. Additionally, accuracy and spatial resolution of optical property mapping can be improved by considering $N$ speckle patterns for each sample (Fig. \ref{fig:overview}). To generate a PSD for each pixel in the image, we first compute a normalized ACF for a 73$\times$73-pixel sliding window, which samples a minimum spatial frequency $0.2mm^{-1}$ due to radially averaging of the ACF about the center pixel. This spatial frequency is commonly used for SFDI \cite{pera2018optical}. The mean ACF ($\overline{ACF}$) is measured as the average of the $N$ windows at each window location. We further improve signal-to-noise ratio by assuming the response to diffuse optical properties is radially-symmetric and averaging the 2D ACF about the radius from each pixel. The same process is performed for a reference phantom with known optical properties. To reconstruct optical properties from the PSD maps, we use a 2-D lookup table with a DC and AC response parameter, similar to conventional SFDI. The DC response ($M_{DC}$) is the remitted signal under planar illumination and can be approximated as:

\begin{equation}
M_{DC} = \frac{\sum_{i=1}^{N}I_i}{N}.
\label{eq:dc}
\end{equation}
where $I_i$ represents raw speckle images. We subtracted the estimated $M_{DC}$ from each speckle image so that the calculated ACF had a zero mean. We define the AC response ($M_{AC}$) as the sum of the frequency response corresponding to the first two points of the PSD curve ($0.1mm^{-1} \leq f \leq 0.5mm^{-1}$, where $f$ stands for spatial frequency). Similar to \cite{cuccia_quantitation_2009}, the DC and AC reflectance ($R_d$) can then be calculated using:

\begin{equation}
R_d = \frac{M}{M_{ref}} \cdot R_{d,ref,pred},
\label{eq:Rd}
\end{equation}
where $M_{ref}$ is the response parameter of the reference phantom, and $R_{d,ref,pred}$ the reflectance value predicted by White Monte Carlo models for the reference. Optical properties are then fit using a lookup table that relates ($\mu_{a}$, $\mu'_{s}$) to ($R_{d,DC}$, $R_{d,AC}$).

\begin{figure}[b]
\centering
\includegraphics[width=\linewidth]{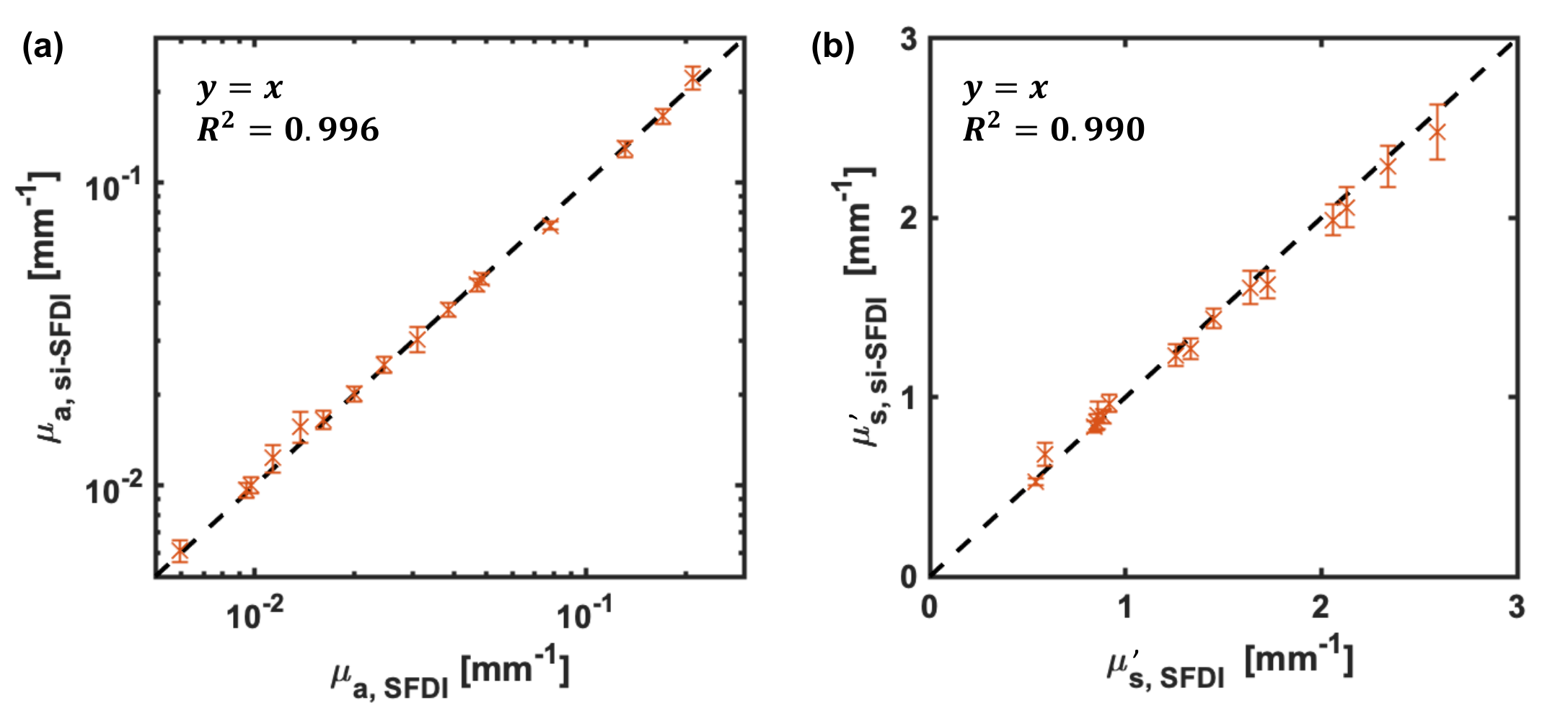}
\caption{Optical property measurements from si-SFDI on 16 homogenous tissue phantoms. (a) si-SFDI measurements of absorption and (b) si-SFDI measurements of reduced scattering versus conventional SFDI ground truth. }
\label{fig:phantoms}
\end{figure}

\begin{figure}[tb]
\centering
\includegraphics[width=\linewidth]{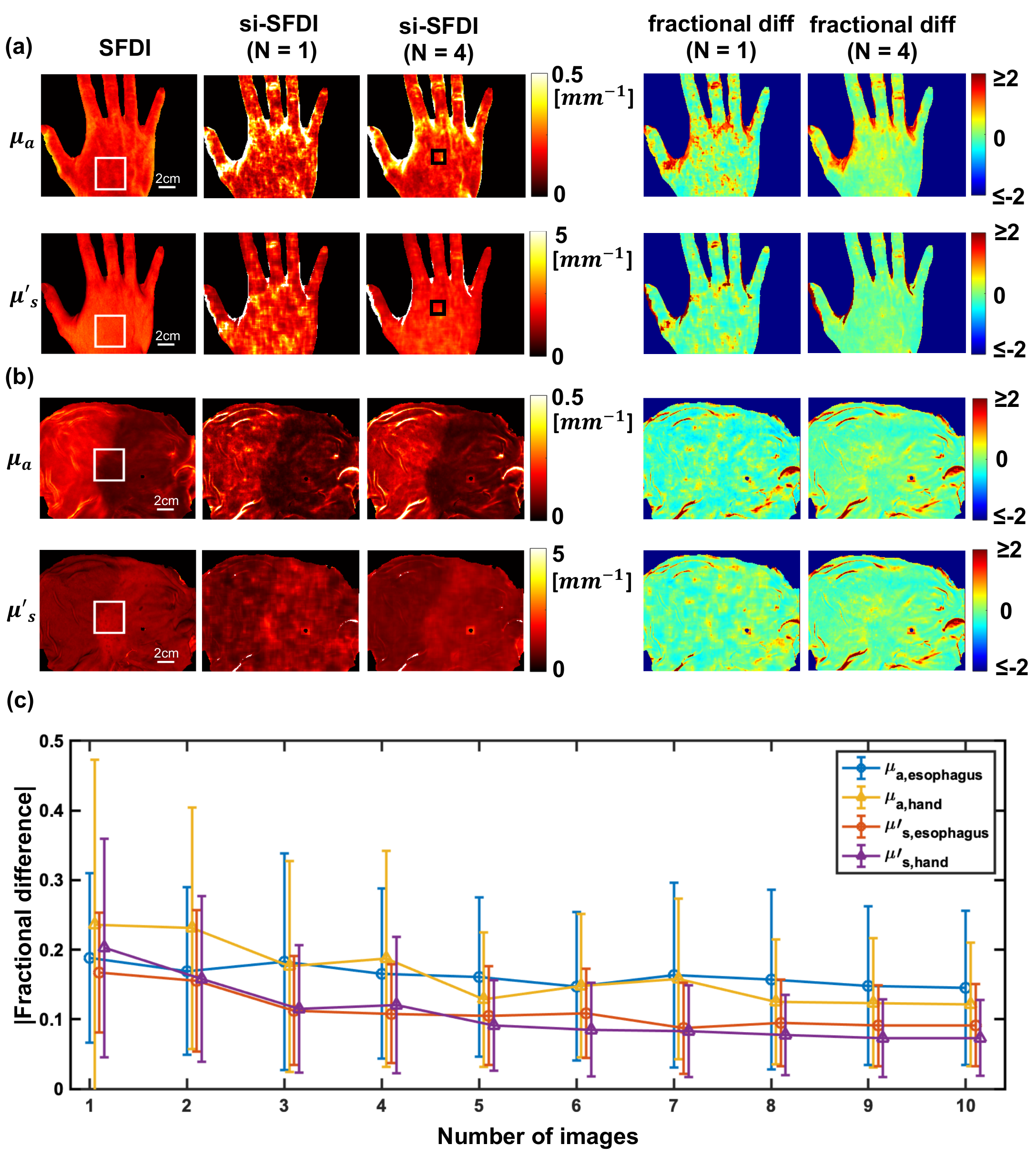}
\caption{Comparison of 1- and 4-image si-SFDI optical property estimates with SFDI ground truth in (a) \textit{in vivo} human hand, and (b) \textit{ex vivo} pig gastroesophageal junction. $N$ indicates the number of input patterns. (c) Absolute percentage difference as a function of number of speckle patterns used in the si-SFDI calculation for the ROI indicated by the white boxes in (a) and (b). Error bars are standard deviations. Black boxes in (a) highlight a scar that did not produce contrast in SFDI.}
\label{fig:results}
\end{figure}

We first validated si-SFDI on homogeneous samples by imaging 16 tissue-mimicking phantoms with unique combinations of optical properties. The phantoms were fabricated by polydimethylsiloxane (PDMS) with titanium dioxide and India ink as scattering and absorption agents, respectively. For each phantom, we compared optical properties estimated by si-SFDI to those computed by conventional SFDI in a central 150$\times$150 pixel region of interest (ROI). For phantoms, the mismatch in illumination wavelengths (520nm for si-SFDI and 526nm for SFDI) was corrected, which is discussed in more detail in Supplement 1. With four input images ($N=4$), si-SFDI results are plotted against SFDI ground truth in Fig. \ref{fig:phantoms}. We found that the si-SFDI predicted optical properties fit a $y=x$ ground truth curve with an $R^2$ value of 0.996 for absorption and 0.991 for reduced scattering. The variance of the si-SFDI measurements increases for larger $\mu'_{s}$, likely due to decreased blurring of the speckle patterns, making the measurements more sensitive to local speckle grains. Analyzing the same ROI across all 16 phantoms, we observed an average pixel error of 6.9\% for $\mu_{a}$ and 5.7\% for $\mu'_{s}$ when using four speckle patterns ($N=4$). Using a single speckle pattern, the average pixel errors increased to 12.9\% for $\mu_{a}$ and 11.0\% for $\mu'_{s}$. Optical property measurement statistics for each phantom pair estimated by 1-image and 4-image si-SFDI are shown in Fig. S1, Supplement 1.

To assess the accuracy of si-SFDI in heterogeneous samples with irregular surface topography, we tested the technique on \textit{in vivo} human hand and \textit{ex vivo} swine gastroesophageal junction (Fig. \ref{fig:results}(a), (b)). Side-by-side comparisons of 4-image si-SFDI with ground truth SFDI show that for flat regions, there is excellent agreement over the whole field of view. si-SFDI recorded a sharp change in optical properties between the stomach and esophagus regions of the swine sample, in agreement with SFDI, and representative of the expected change in tissue type on either side of the junction.

To explore the effect of the number of speckle patterns on the accuracy of optical property calculations, we analyzed a relatively-flat region on the back of the hand and at the gastroesophageal junction (250$\times$250 pixels, highlighted by the white boxes in Fig. \ref{fig:results}(a) and (b)) and plotted the error with varying numbers of images (Fig. \ref{fig:results}(c)). This error was calculated as the absolute percentage difference. The errors for both absorption and scattering measurements decrease significantly when additional speckle patterns are included, but beyond $N=4$ the improvements become small--the decrease in errors from $N=4$ to $N=10$ images is only approximately 2\%. Analyzing additional speckle patterns has the effect of both increasing signal to noise and improving the spatial resolution of optical property mapping, since additional patterns improve the sampling of the local ACF in our window-based approach. Single-image si-SFDI results ($N=1$) are also shown in Fig. \ref{fig:results}(a) and (b). As expected, the results show fair agreement with SFDI, however, with more noise and image artifacts than $N=4$. To further investigate the accuracy and image quality of si-SFDI applied to heterogeneous samples, we calculate the structured similarity (SSIM) index for the pig esophagus and human hand sample (Table \ref{tab:ssim}). Overall, si-SFDI demonstrates high SSIM scores compared to profile-corrected SFDI as reference. On average, si-SFDI with 4 images ($N=4$) achieves 6.3\% higher SSIM than $N=1$, and this improvement becomes 8.2\% with $N=10$. 

For both hand and esophagus samples (Fig. \ref{fig:results}), we observe a larger discrepancy between si-SFDI and ground truth absorption than scattering measurements. This may be partially due to the mismatch in wavelengths used (520nm and 526nm for si-SFDI and SFDI, respectively). For example, the extinction coefficients of hemoglobin are different at these two wavelengths, with a ratio of approximately 1 to 1.2 \cite{prahl1998tabulated}. This mismatch could also contribute to the differences in scattering, but likely to a lesser extent \cite{tseng2009chromophore}. Moreover, in areas of the sample with irregular surface topography, we found larger errors in absorption than scattering estimates. This is expected because absorption depends mostly on $M_{DC}$, while scattering depends mostly on $M_{AC}$, which is less affected by surface angle variations. Additionally, the difference in illumination angle for the laser diode compared to the SFDI projector is expected to contribute to reconstruction errors. These effects are further explored and discussed in Supplement 1. Compared to profile-corrected SFDI, the pixel error rate of si-SFDI is below 20\% when the surface is flat (<10 degrees) and increases with larger angles. 

\begin{table}
\caption{SSIM of pig esophagus and human hand optical properties obtained by si-SFDI. SFDI is used as reference.} 
\label{tab:ssim}
\begin{center}       
\begin{tabular}{|p{2.7cm}|c|c|c|c|}
\hline
 \multirow{2}{2.7cm}{Number of speckle patterns ($N$)} & \multicolumn{2}{c|}{Pig esophagus} & \multicolumn{2}{c|}{Human hand} \\
\cline{2-5}
 & $\mu_{a}$ & $\mu'_{s}$ & $\mu_{a}$ & $\mu'_{s}$ \\
\cline{2-5}
\hline
$N=1$ & 0.790 & 0.870 & 0.777 & 0.817\\
\hline
$N=4$ & 0.839 & 0.901 & 0.846 & 0.871\\
\hline
$N=10$ & 0.855 & 0.912 & 0.874 & 0.893\\ 
\hline
\end{tabular}
\end{center}
\vspace*{-5mm}
\end{table}

We observed a small region with large errors on the hand sample (highlighted by the black boxes in Fig. \ref{fig:results}(a)). From inspection of the hand, we found a scar in this region that does not generate contrast in conventional SFDI but was captured with si-SFDI. This difference in contrast may be due to the si-SFDI incorporating higher spatial frequencies than SFDI, which would lead to a shallower sampling of tissue optical properties \cite{cuccia_quantitation_2009} and also contributions from sub-diffuse scattering parameters. A recent study similarly uncovered a scar when using high spatial frequency structured illumination \cite{kanick_sub-diffusive_2014}. We note that even at the high spatial frequencies analyzed in si-SFDI, we did not observe a change in optical property estimation when blood flow was occluded (Supplement 1).

There are several areas for improvement of the si-SFDI technique. First, multiple speckle patterns are required for the most accurate optical property estimate. Although this can be achieved by using a rotating diffuser or phase randomization, it inevitably limits the imaging speed. However, instead of averaging multiple speckle patterns to approximate the DC response, one can use a laser speckle reducer to approximate planar illumination and accurately estimate $M_{DC}$ from a single image. In this scenario, only two images would be required. Moreover, if the average illumination power of these two images is different and known, sensitivity to ambient lighting can be reduced. Alternatively, planar illumination may be approximated from coherent illumination using a content-aware deep learning approach \cite{bobrow2019deeplsr}, allowing accurate single-frame si-SFDI. Second, the processing speed of si-SFDI is slow due to window-based computations. It takes 2 seconds to compute a 1040$\times$1392-pixel optical property map from a single speckle pattern ($N=1$) using a 4-core 3.6 GHz processor running the algorithm in MATLAB (R2020a, MathWorks). However, because each window computation is independent, this computation is highly parallelizable and can be accelerated by using a graphics processing unit. Third, for many clinical applications, the topography of the sample will need to be estimated and accounted for. This may be accomplished in simple setups with recent monocular depth estimation techniques \cite{mahmood2018deep,chen2018rethinking}. 

To summarize, we demonstrate a wide-field, projector-free technique for non-contact optical property mapping using laser speckle patterns as structured illumination. When applied to both homogeneous phantoms and heterogeneous biological tissues, si-SFDI accurately measures scattering and absorption parameters. With further optimization, this technique can facilitate the translation of quantitative optical property mapping in more clinically-friendly systems including endoscopes.

\section*{Funding Information}
Supported in part with funding from an NIH NIBIB Trailblazer Award (R21 EB024700).

\section*{Disclosures} 
The authors declare no conflicts of interest.

\clearpage

\bibliographystyle{unsrt}  

\bibliography{ref.bib}



\end{document}